\documentclass[twocolumn]{webofc}

\usepackage[varg]{txfonts}   
\usepackage{hyperref}
\usepackage{url}
\usepackage{subcaption}
\hypersetup{colorlinks=true,citecolor=blue,urlcolor=blue,linkcolor=blue}
\usepackage[labelfont=bf,format=plain]{caption}
\usepackage{threeparttable}

\usepackage{atbegshi}
\AtBeginShipout{%
  \ifnum\value{page}=5
    \AtBeginShipoutDiscard
  \fi
}

\begin{document}
\title{Fast-reactor neutron sources in evaluated nuclear data library validation}
%
%

\author{\firstname{Aaron} \lastname{Hurst}\inst{1,2}\fnsep\thanks{\email{amhurst@berkeley.edu}} \and
  \firstname{Emanuel} \lastname{Chimanski}\inst{3} \and
  \firstname{Toshihiko} \lastname{Kawano}\inst{4} \and
  \firstname{David} \lastname{Brown}\inst{3} \and
  \firstname{David} \lastname{Matters}\inst{2} \and
  \firstname{Lee} \lastname{Bernstein}\inst{1,2} 
}

\institute{Department of Nuclear Engineering, University of California, Berkeley, California 94720, USA
  \and
  Nuclear Science Division, Lawrence Berkeley National Laboratory, California 94720, USA
  \and
  National Nuclear Data Center, Brookhaven National Laboratory, Upton, New York 11973, USA
  \and
  Theoretical Division, Los Alamos National Laboratory, Los Alamos, New Mexico 87545, USA
}

\abstract{
  Two different neutron sources, based on $^{235}$U-fission neutrons with different average energies, provide integral benchmark data for validation of $\gamma$-ray production data in the evaluated nuclear data libraries corresponding to fast-neutron-induced inelastic-neutron scattering reactions.  Firstly, we consider the IRT-M Research Reactor, formerly located at the Nuclear Research Institute just outside of Baghdad, Iraq, to demonstrate the validation methodology using the associated $\gamma$-ray data in the Evaluated Nuclear Data File, version VIII.0 (ENDF/B-VIII.0), for several $\gamma$-ray transitions over a wide range of nuclides including $^{28}$Si, $^{32}$S, $^{56}$Fe, and $^{186}$W.  Using the characterized neutron flux of the Baghdad IRT-M Reactor, we find flux-weighted cross-sections deduced using the ENDF/B-VIII.0 $\gamma$-ray data to be in good agreement with the integral measurements performed at the Baghdad Research Reactor in addition to the corresponding results of different reaction-model calculations, {\tt CoH$_{3}$} and {\tt EMPIRE}.  Given the excitation thresholds for the $\gamma$-ray transitions involved in this investigation, these observations lend further support to the characterization of the IRT-M flux in the fast-neutron energy region $0.862 \leq E_{n} \leq 5.0$ MeV.  The additional detail devoted to the IRT-M source reflects the broader scope of the validation work carried out at that facility.  A second neutron source considered for this validation work is the Forschungsreaktor M{\"u}nich (FRM-II), Garching, Germany.  Again, the flux-weighted $\gamma$-ray data from ENDF/B-VIII.0 for $^{56}$Fe compare well to the integral FRM-II measurement and reaction-model calculations.

}
\maketitle
\section{Introduction}
\label{intro}

Validation of $\gamma$-ray production data from fast-neutron-induced inelastic-scattering reactions can provide valuable benchmark integral data for the evaluated nuclear data libraries, such as the Evaluated Nuclear Data File, version VIII (ENDF/B-VIII.0) \cite{brown:18}.  In this work we have considered two $^{235}$U-fission-based neutron sources with very different average energies for this purpose: the IRT-M Baghdad Research Reactor \cite{demidov:78,hurst:21} and the Forschungsreaktor M{\"u}nchen II (FRM-II) reactor \cite{ilic:20}.

Regarding the first of these neutron sources, in a perspective on current nuclear data needs for applications it is claimed that the flux of the Baghdad Reactor \textit{``was not well-characterized''} \cite{kolos:22}.  This assertion may be related to a statement made in the context of experimental benchmarks at the WANDA 2021 meeting \cite{lewis:21}.  In these proceedings, we provide evidence that the flux developed in Ref.~\cite{hurst:21} is well characterized and suitable for use as an integral benchmark, demonstrating that the functional form of the flux, from a covariance analysis of the measured $^{56}$Fe data in a $\chi^{2}$ minimization, can not only accurately reproduce the measured experimental data, but also the corresponding flux-weighted $\gamma$-ray cross-section data from ENDF/B-VIII.0 \cite{brown:18}.  Lending further support to the validity of the derived flux \cite{hurst:21}, we also demonstrate consistency between integral measurements carried out at the Baghdad Research Reactor and flux-weighted cross sections deduced from the ENDF/B-VIII.0 library and reaction-model calculations for additional nuclides including $^{28}$Si, $^{32}$S, and $^{186}$W.  Flux-weighted quantities were deduced for several transitions in these nuclides that, together with the $^{56}$Fe data, collectively span a fast-neutron energy region of $0.862 \leq E_{n} \leq 5.0$~MeV, implying that the flux is well characterized over this energy region and suitable for use as an integral benchmark.

An ongoing campaign to measure and develop a catalogue of $(n,n'\gamma)$ data is also currently underway at the FRM-II.  Accordingly, this provides an additional neutron source, though with a different average neutron energy, that we can exploit to test our methodology for the validation of integral benchmark data in the fast neutron-energy region.  In this work, we also compare and demonstrate consistency between the integral FRM-II measurement for $^{56}$Fe \cite{ilic:20} with flux-weighted cross sections deduced from data in the ENDF/B-VIII.0 library as well cross sections based on reaction-model calculations.

\section{Neutron sources for integral benchmarks}
\label{sec-sources}

\begin{figure}[h]
  \centering
  \includegraphics[width=0.98\linewidth,clip]{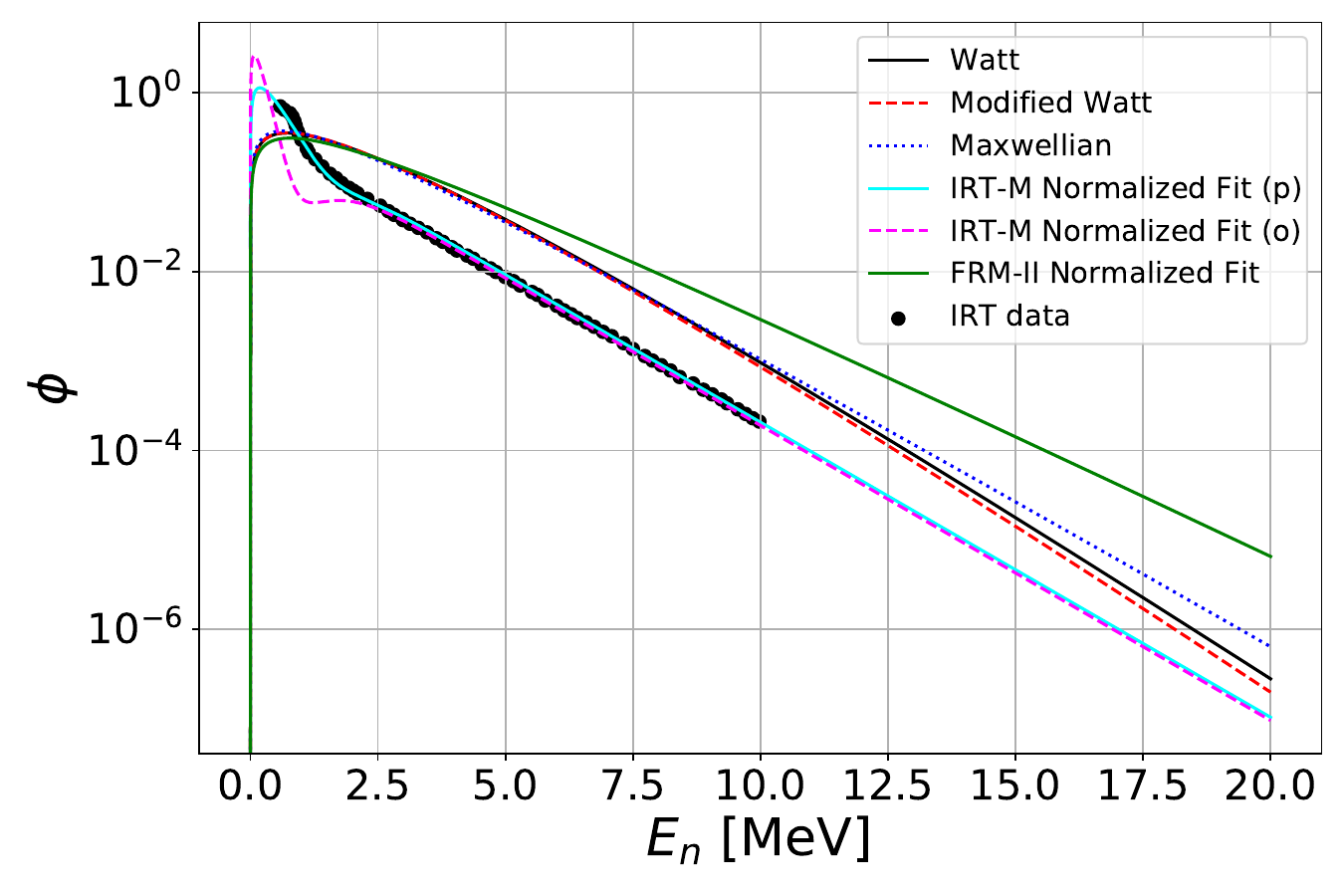}
  \caption{The normalized reported flux at the IRT-M Baghdad Research Reactor (black circles) \cite{demidov:78} presented in comparison to parametrizations described in Refs.~\cite{hurst:21, hurst:24} alongside standard distributions from $^{235}$U fission neutrons: Watt, Modified Watt, and Maxwellian (see Ref.~\cite{hurst:24}).  The normalized FRM-II flux \cite{ilic:20} is also shown.}
  \label{fig:1}
\end{figure}

Differential $\gamma$-ray production data from ($n,n'$) reaction channels can be used for validation purposes when combined with a well-characterized neutron flux over a suitable range of incident neutron energies.  These data may be taken from, or more precisely, derived from evaluated libraries such as ENDF/B-VIII.0 \cite{brown:18}.  Alternatively, reaction models such as \texttt{CoH$_{3}$} \cite{kawano:19} and \texttt{EMPIRE} \cite{herman:07} can be tuned to reproduce the expected $\gamma$-ray production spectrum from ($n,n'$).  The $\gamma$-ray production data ($\sigma_{\gamma}$) as a function of incident-neutron energy ($E_{n}$) can then be used to provide a flux-weighted average of the cross section for a given transition ($E_{\gamma}$) when convolved with the appropriately characterized neutron flux $\phi(E_{n})$ in accordance with
\begin{equation}
  \label{eq:1}
  \langle \sigma_{\gamma}(E_{\gamma}) \rangle = \frac{\int_{E_{\text{th}}}^{E_{n}=E} \phi(E_{n}) \sigma_{\gamma}(E_{n}) W_{\gamma}(\theta=90^{\circ};E_{n})dE_{n}}{\int_{0}^{+\infty} \phi(E_{n})dE_{n}},
\end{equation}
where the lower limit in the numerator $E_{\text{th}}$ is the threshold energy of the $\gamma$-ray producing ($n,n'$) channel, and the upper limit $E$ is where the contribution to the integral becomes negligible for the neutron source under consideration.  When possible, the angular-distribution correction factor $W_{\gamma}$ should be folded into the convolution process.  For further details on the determination of $\langle \sigma_{\gamma}(E_{\gamma}) \rangle$ see Refs.~\cite{hurst:21, hurst:24}.  Integral measurements of the $\gamma$-ray production data obtained from the Baghdad Reactor and FRM-II can then be compared to the flux-weighted quantities deduced using Eq.~(\ref{eq:1}).  An overview of these results is given below for each source, in turn.

\subsection{Baghdad Research Reactor}
\label{sec-baghdad}

The functional form of the neutron flux $\phi(E_{n})$ for the Baghdad Reactor is derived in Refs.~\cite{hurst:21, hurst:24} and shown in Fig.~\ref{fig:1}.  In Ref.~\cite{hurst:24} it is explicitly shown that these parameterized distributions satisfy the normalization condition allowing for an appropriate comparison with standard normalized distributions associated with fission neutrons from $^{235}$U: Watt; modified Watt; Maxwellian \cite{hurst:24}.  The normalized distributions are all also shown in Fig.~\ref{fig:1} where it is apparent that the standard $^{235}$U fission-based distributions, all of which have an average neutron energy of $\sim 2$~MeV, have significantly different profiles to the parameterized distributions of the Baghdad Reactor with average neutron energies of $<1$~MeV \cite{hurst:24}.  The beam profile at the target station of the fast-filtered beamline of the Baghdad Reactor is, thus, expected to exhibit significantly different characteristics to that of a pure unmoderated/unfiltered Watt- or Maxwellian-type spectrum.

\begin{figure*}[t]
  \centering
  \begin{subfigure}[b]{0.33\textwidth}
    \includegraphics[width=\linewidth,clip]{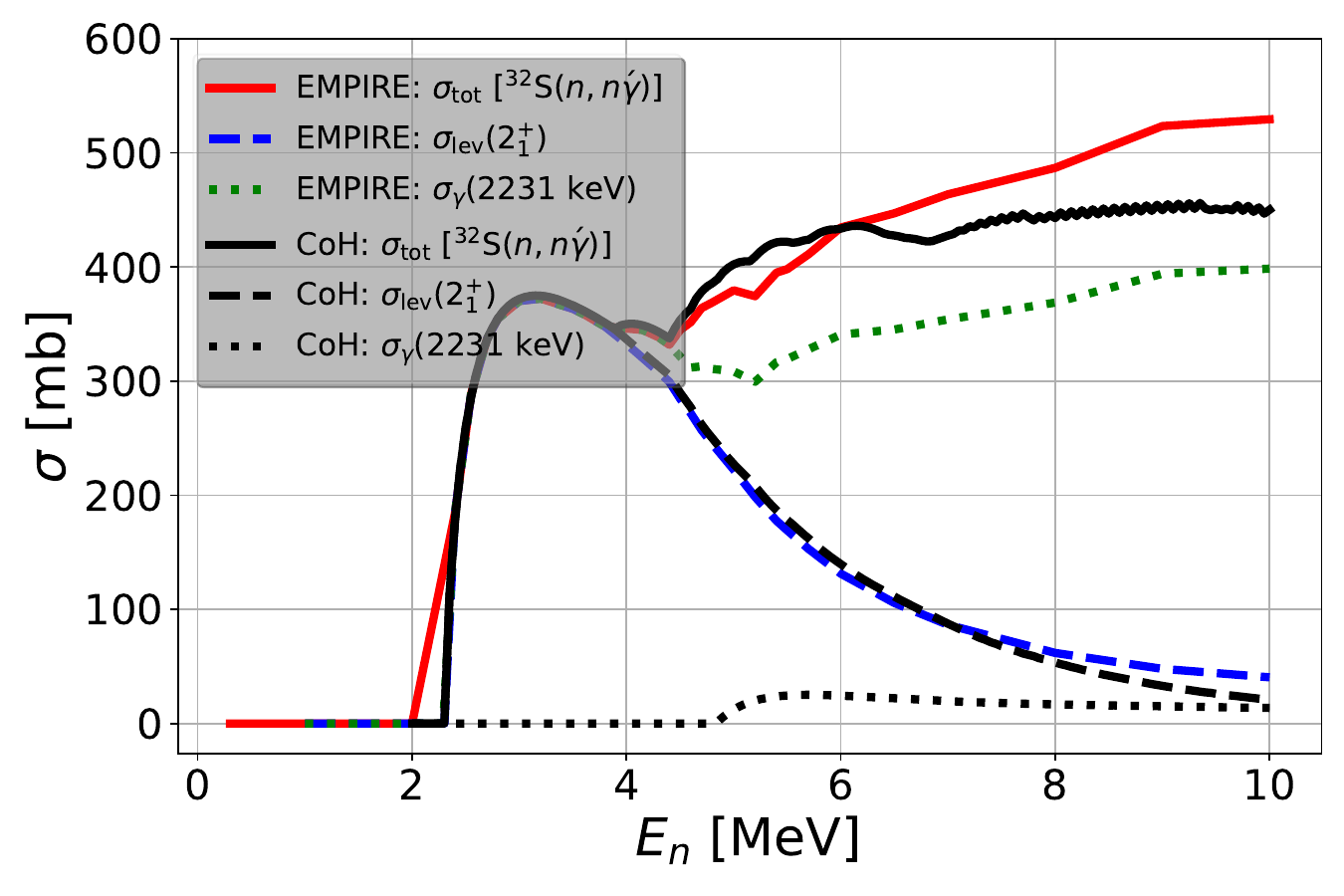}
    \caption{}
  \end{subfigure}
  \begin{subfigure}[b]{0.33\textwidth}
    \includegraphics[width=\linewidth,clip]{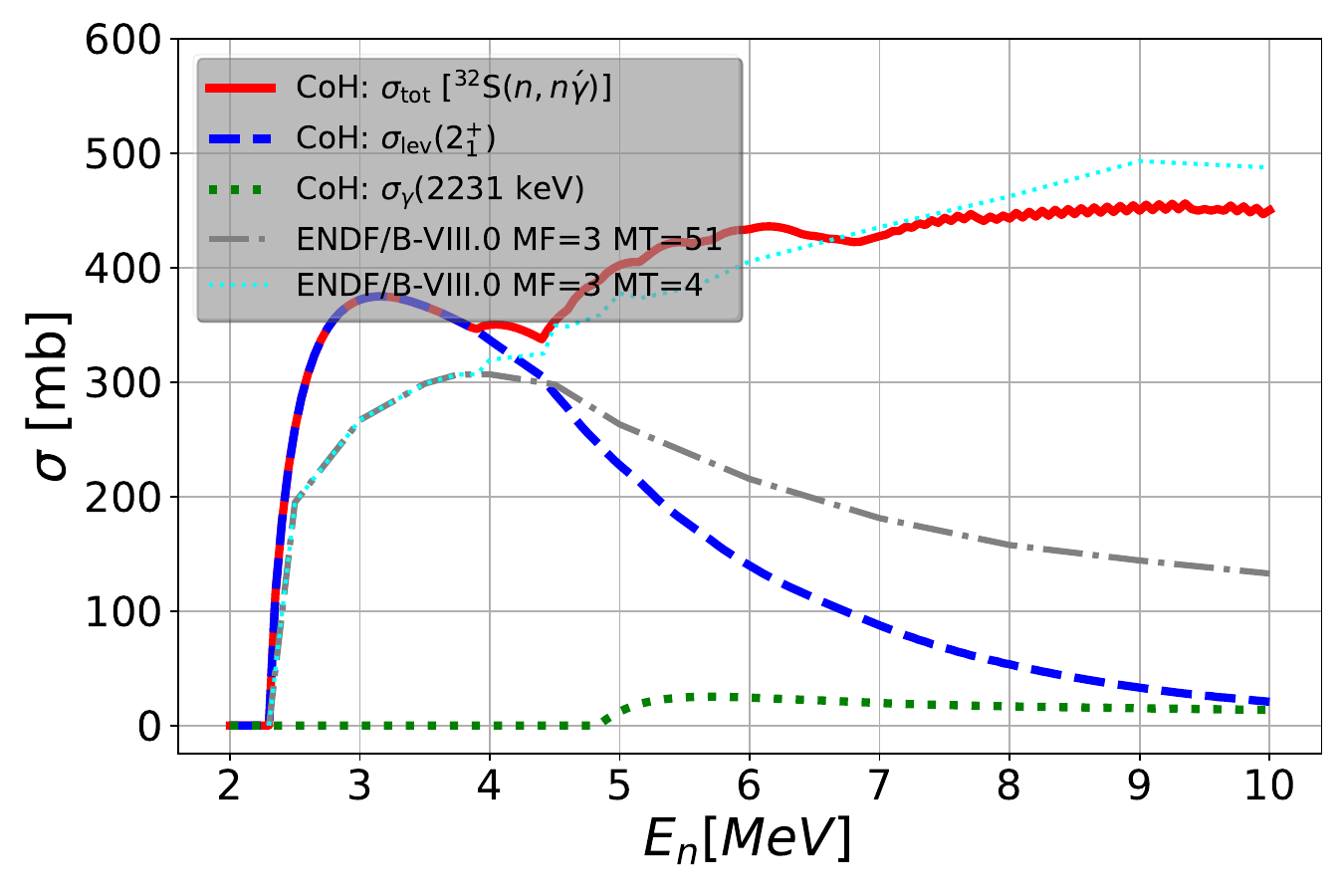}
    \caption{}
  \end{subfigure}
  \begin{subfigure}[b]{0.33\textwidth}
    \includegraphics[width=\linewidth,clip]{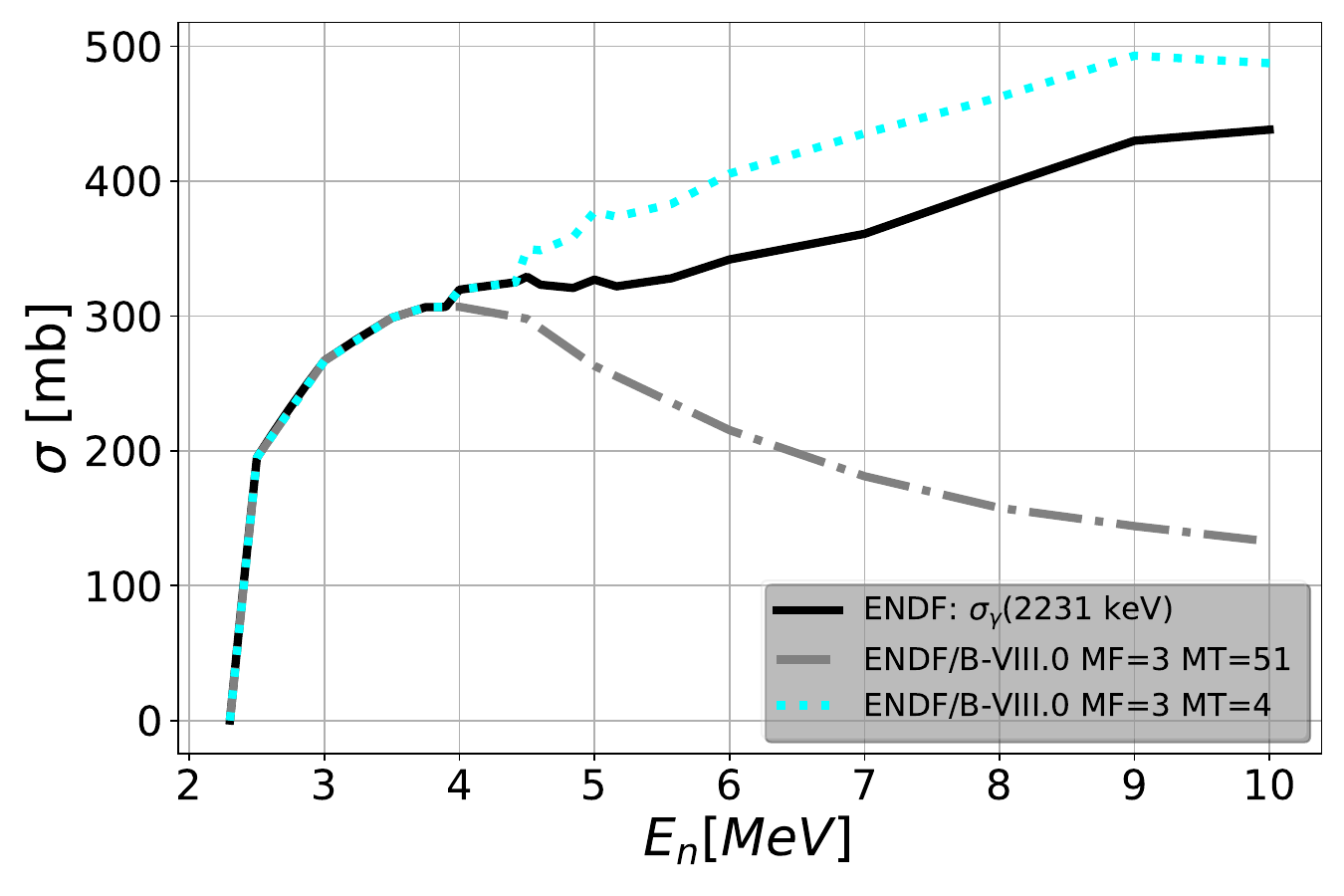}
    \caption{}
  \end{subfigure}
  
  \caption{(a) Calculated cross sections for $^{32}$S($n,n'\gamma$) obtained using \texttt{CoH$_{3}$} and \texttt{EMPIRE}.  (b) The ENDF/B-VIII.0 data for $^{32}$S($n,n'\gamma$) compared to reaction-model results: \texttt{MF3 MT51} cf. $\sigma_{\text{lev}}(2^{+}_{1})$; \texttt{MF3 MT4} cf. $\sigma_{\text{tot}}$[$^{32}$S($n,n'\gamma$)].  (c) The $\gamma$-ray production data (solid black line) derived from ENDF/B-VIII.0 (discussed in Ref.~\cite{hurst:24}) for the 2231-keV $2_{1}^{+} \rightarrow 0^{+}_{\text{gs}}$ in $^{32}$S.}
  \label{fig:2}
\end{figure*}

In Ref.~\cite{hurst:21} it is shown that an appropriate treatment and corresponding parametrization of the reported flux data in Ref.~\cite{demidov:78}, whereupon the low-energy region $E_{n} \ll 1.5$~MeV is treated as a Maxwellian distribution while neutrons of higher $E_{n}$ follow a simple exponential (and, thus, verifying the claim of Ref.~\cite{demidov:78}), can already adequately reproduce the measured $\gamma$-ray intensity data for $^{56}$Fe; the nucleus which defines the absolute $\gamma$-ray intensity normalization of the entire atlas.  The resulting curve is labeled as the parameterized fit in Fig.~\ref{fig:1} (cyan curve).  A temperature adjustment of the neutron spectrum in an iterative $\chi^{2}$ minimization improved agreement with the measured data, resulting in the curve labeled as the optimized fit in Fig.~\ref{fig:1} (magenta curve).  The minimization procedure is based on a covariance analysis of the three strongest transitions observed in $^{56}$Fe($n,n'\gamma$) and is explained in detail in Ref.~\cite{hurst:21}.  A consistent set of flux-weighted partial $\gamma$-ray production cross sections are obtained for the three strongest transitions in $^{56}$Fe \cite{hurst:21, hurst:24} whether using data from the ENDF/B-VIII.0 library \cite{brown:18} or reaction-model calculations using \texttt{CoH$_{3}$} \cite{kawano:19}, and also found to be in agreement with the measured integral $^{56}$Fe data.

\begin{table}[b!]
  \caption{Flux-weighted quantities for the $^{32}$S($n,n'\gamma$) reaction corresponding to the \texttt{CoH$_{3}$} and \texttt{EMPIRE} reaction-model calculations and values based on ENDF/B-VIII.0.  The integral $\gamma$-ray production cross section for the 2231-keV $2^{+}_{1} \rightarrow 0^{+}_{\text{gs}}$ transition deduced from the Baghdad Reactor measurement is also listed.  All cross sections are given in units of [mb].}
  \label{tab:1}
  \begin{threeparttable}
    \centering
    \begin{tabular}{lllll}\hline
      Method & $\langle \sigma_{\gamma} \rangle$ & $\langle \sigma_{\rm tot} \rangle$ & $\langle \sigma_{\rm lev} \rangle$ & $\frac{\langle \sigma_{\rm lev} \rangle}{\langle \sigma_{\rm tot} \rangle}$\\
      \hline
      CoH & 26.57 & 27.91 & 23.95 & 0.858 \\
      EMPIRE & 26.49 & 28.80 & 23.78 & 0.890 \\
      ENDF\tnote{a} & 21.67 & 22.76 & 19.98 & 0.878 \\
      ENDF\tnote{b} & 21.79 & $-$ & $-$ & $-$ \\
      Baghdad Atlas & \textbf{25.0(60)} & $-$ & $-$ & $-$ \\
      \hline
    \end{tabular}
    \begin{tablenotes}
    \item[a] Estimated result.
    \item[b] Derived result.
    \end{tablenotes}
  \end{threeparttable}
\end{table}

Following on from the initial work on $^{56}$Fe, the validation methodology has been applied to additional integral measurements carried out at the Baghdad Reactor.  For example, Table~\ref{tab:1} lists the flux-weighted $\gamma$-ray production cross sections for the $2^{+}_{1} \rightarrow 0^{+}_{\text{gs}}$ transition in $^{32}$S at 2231~keV.  These values were deduced using calculated $\gamma$-ray production spectra from the reaction models embodied within \texttt{CoH$_{3}$} \cite{kawano:19} and \texttt{EMPIRE} \cite{herman:07} (Fig.~\ref{fig:2}(a)), in addition to data from the ENDF/B-VIII.0 library (Fig.~\ref{fig:2}(b)).  All $\langle \sigma_{\gamma} \rangle$ results are statistically consistent with the integral measurement of the Baghdad Reactor that is determined on an absolute scale relative to the 847-keV $2^{+}_{1} \rightarrow 0^{+}_{\text{gs}}$ transition in $^{56}$Fe.  Also listed in Table~\ref{tab:1} are the flux-weighted cross sections for the total $^{32}$S$(n,n'\gamma)$ reaction cross section $\langle \sigma_{\text{tot}} \rangle$ and the direct excitation function for the first $2^{+}$ state in $^{32}$S $\langle \sigma_{\text{lev}} \rangle$.  These results were calculated in an analogous manner to $\langle \sigma_{\gamma} \rangle$ using Eq.~(\ref{eq:1}).  As explained in Refs.~\cite{hurst:21, hurst:24}, $\sigma_{\gamma}$ can not be extracted directly from the ENDF libraries in their current construct, unlike $\sigma_{\text{tot}}$ and $\sigma_{\text{lev}}$.  However, provided that the ratio $\langle \sigma_{\text{lev}} \rangle/\langle \sigma_{\text{tot}} \rangle$ from a given reaction model is approximately consistent to the corresponding ratio from the ENDF library, which holds in the case of $^{32}$S($n,n'\gamma$) as shown by the results in the final column of Table~\ref{tab:1} where all ratios agree to within $\sim 1\%$, we can estimate $\sigma_{\gamma}$ according to the methodology outlined in Refs.~\cite{hurst:21, hurst:24}.  Alternatively, in the more recent development described in Ref.~\cite{hurst:24}, a new procedure has been proposed to derive $\sigma_{\gamma}$ as a function of $E_{n}$ after accounting for direct excitation, feeding from higher-lying discrete levels, and quasicontinuum contributions (Fig.~\ref{fig:2}(c)).  Numerically, the \textit{estimated} and \textit{derived} results obtained using ENDF/B-VIII.0 data are very close and both in statistical agreement with the integral measurement of the Baghdad Reactor.  At lower incident-neutron energies there are differences between the reaction-model results and the ENDF/B-VIII.0 library which likely explains the difference in overall flux-weighted results here.  Ultimately, however, all results are consistent with the integral data for $^{32}$S($n,n'\gamma$).

\begin{figure}[b]
  \centering
  \includegraphics[width=0.98\linewidth,clip]{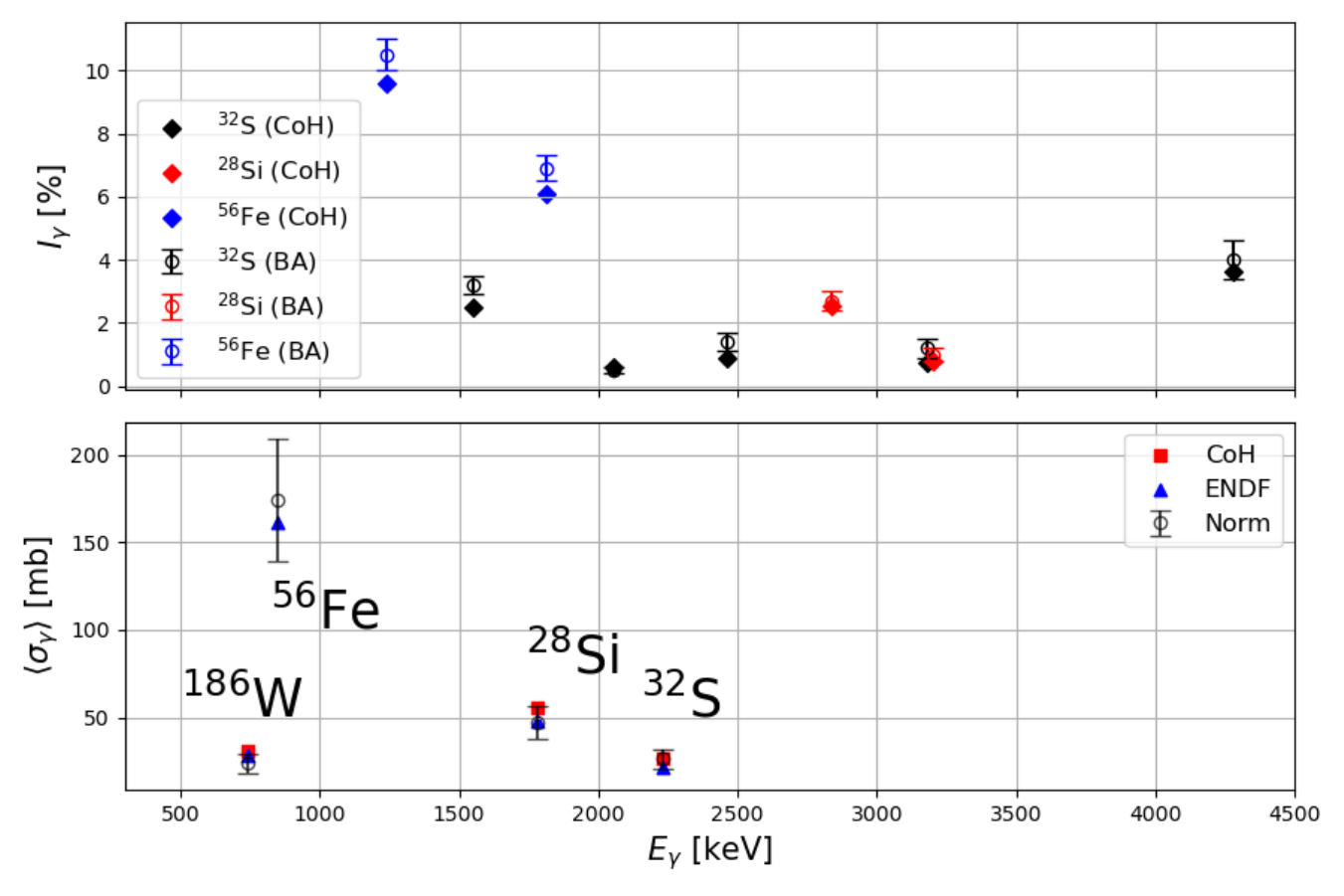}
  \caption{Lower panel: Plot showing integral cross sections measured at the Baghdad Reactor for the normalization $\gamma$-ray transitions (see text) observed in a given nuclide compared to flux-weighted results obtained using the \texttt{CoH$_{3}$} reaction model and data from ENDF/B-VIII.0.  Upper panel: Relative intensities of integral measurements for weaker transitions in the corresponding nuclide datasets in comparison to the \texttt{CoH$_{3}$} reaction-model calculations.}
  \label{fig:3}
\end{figure}

The results of additional $\gamma$-ray lines in $^{28}$Si, $^{32}$S, $^{56}$Fe, and $^{186}$W that have been investigated as part of this validation task are summarized in Fig.~\ref{fig:3}; for consistency all results neglect the $W_{\gamma}$ correction.  Each dataset of the Baghdad Atlas \cite{demidov:78,hurst:21} has a normalization transition, usually the strongest and defined with a relative intensity of 100\%, which in turn is defined relative to the $2^{+}_{1} \rightarrow 0^{+}_{\text{gs}}$ transition in $^{56}$Fe allowing for absolute determination of all integral $\gamma$-ray intensities through a two-stage normalization process.  The integral cross sections for the normalization transitions associated with all aforementioned nuclei are shown in the lower panel of Fig.~\ref{fig:3} in comparison to flux-weighted cross sections based on reaction model calculations and data from the ENDF/B-VIII.0 library.  The upper panel compares the $\gamma$-ray intensities of weaker lines from the integral measurements with calculated results from \texttt{CoH$_{3}$}.  Generally, Fig.~\ref{fig:3} shows good agreement across the board for several nuclei of different mass with transitions associated with level-energy excitation thresholds ranging from around 862~keV to 5~MeV.  These observations reinforce the findings of Ref.~\cite{hurst:21} and imply the flux is well characterized over the neutron energy region $0.862 \leq E_{n} \leq 5.0$~MeV.

\subsection{Forschungsreaktor M{\"u}nchen II}
\label{sec-frm2}

A concerted effort is underway at the FRM-II to develop a new catalogue of ($n,n'\gamma$) data, similar to that of the Baghdad Atlas \cite{demidov:78, hurst:21}, based on a neutron source with a higher average energy of 2.3~MeV and a neutron flux of $1.4 \times 10^{8}$ cm$^{-2}$s$^{-1}$ \cite{ilic:20} using the Fast Neutron Gamma Spectroscopy (FaNGaS) instrument \cite{randriamala:16}.  In a preliminary assessment of the flux used to measure $^{56}$Fe($n,n'\gamma$) \cite{ilic:20} at the FRM-II, which qualitatively has a profile similar to that of a Maxwellian distribution as shown by its normalized distribution in Fig.~\ref{fig:1}, we were able use a simple Maxwellian to approximately describe the flux distribution and reproduce the expected average neutron energy of 2.3~MeV.  Using the same validation methodology and approach outlined in Sect.~\ref{sec-baghdad}, we present the flux-weighted cross-section results from \texttt{CoH$_{3}$} and ENDF/B-VIII.0 in comparison to the integral cross section for the 847-keV $2^{+}_{1} \rightarrow 0^{+}_{\text{gs}}$ transition in $^{56}$Fe measured at the FRM-II in Table~\ref{tab:2}.  Here, we list the $\gamma$-ray production cross section results both with and without the $W_{\gamma}$ correction factor folded in.  As before with $^{32}$S, we also find the ratio $\langle \sigma_{\text{lev}} \rangle/\langle \sigma_{\text{tot}} \rangle$ based on the \texttt{CoH$_{3}$} calculations to be in close agreement with the same ratio from the ENDF/B-VIII.0 library.  Moreover, the $\langle \sigma_{\gamma} \rangle$ results from \texttt{CoH$_{3}$} and ENDF/B-VIII.0 are statistically consistent with the integral measurement of the FRM-II, both with and without the angular-distribution correction factored in, thus, also supporting the adopted validation methodology for this higher-energy neutron source.

\begin{table}[b!]
  \caption{Flux-weighted quantities for the $^{56}$Fe($n,n'\gamma$) reaction corresponding to the \texttt{CoH$_{3}$} reaction-model calculations and values based on ENDF/B-VIII.0.  The integral $\gamma$-ray production cross section for the 847-keV $2^{+}_{1} \rightarrow 0^{+}_{\text{gs}}$ transition deduced from the FRM-II measurement is also listed, both with and without the $W_{\gamma}$ correction folded in.  All cross sections are given in units of [mb].}
  \label{tab:2}
  \begin{threeparttable}
    \centering
    \begin{tabular}{llllll}\hline
      Method & $\langle \sigma_{\gamma} \rangle_{W_{\gamma}}$ & $\langle \sigma_{\gamma} \rangle$ & $\langle \sigma_{\rm tot} \rangle$ & $\langle \sigma_{\rm lev} \rangle$ & $\frac{\langle \sigma_{\rm lev} \rangle}{\langle \sigma_{\rm tot} \rangle}$\\
      \hline
      CoH & 622.5 & 759.4 & 797.3 & 468.8 & 0.59 \\
      ENDF\tnote{a} & 590.7 & 720.7 & 756.6 & 461.3 & 0.61 \\
      ENDF\tnote{b} & 552.6 & 673.9 & $-$ & $-$ & $-$\\
      FRM-II & \textbf{586(41)} & \textbf{715(50)} & $-$ & $-$ & $-$\\
      \hline
    \end{tabular}
    \begin{tablenotes}
    \item[a] Estimated result.
    \item[b] Derived result.
    \end{tablenotes}
  \end{threeparttable}
\end{table}

\section{Summary and Outlook}
\label{sec-summary}

Integral ($n,n'\gamma$) cross-section data measured at the Baghdad Research Reactor and the FRM-II reactor have been used to validate the ENDF/B-VIII.0 libraries in the fast-neutron energy region.  We have applied our validation methodology to several transitions over a range of nuclei of various mass.  In all cases considered, we find the integral measurements to be in good agreement with the corresponding flux-weighted cross sections based on the ENDF/B-VIII.0 library data.  Flux-weighted cross sections from reaction-model calculated $\gamma$-ray production spectra using \texttt{CoH$_{3}$} and \texttt{EMPIRE} also support these findings.

The Baghdad Reactor measurements involve $\gamma$-ray transitions associated with level-excitation energy thresholds from around 862~keV to 5~MeV in different nuclei: $^{28}$Si, $^{32}$S, $^{56}$Fe, and $^{186}$W.  Because of the good agreement between integral and flux-weighted cross sections, these results confirm that the flux developed in Ref.~\cite{hurst:21} is well characterized over the neutron energy range $0.862 \leq E_{n} \leq 5.0$~MeV.

In the future, we intend to investigate additional transitions from different nuclei measured at both the Baghdad Reactor and the FRM-II to further test our validation methodology.  An improved characterization of the FRM-II flux is needed, while additional low-energy data points would be useful to help constrain the low-energy tail of the Baghdad Reactor flux.

\section*{Acknowledgments}
\label{sec-ack}

This work was performed under the auspices of the DOE NNSA by the NNSC under Award Number DE-NA0003996 at UC Berkeley, LBNL under Contract No. DE-AC0205CH11231, BNL under Contract No. DE-AC02-98CH10886, and by LANL under Contract No. 89233218CNA000001.  The authors thank Dr. Eric Mauerhofer for providing the FRM-II neutron flux data.

%
%
%

\end{document}